\documentclass[10pt]{article}
\begin{document}
\parindent=1.0cm 
\begin{center}

{\large \bf Phase Transitions: Summary of Discussion Session II 
of Camerino 2005}
\bigskip
{\large 

{Dennis Bonatsos}

{Institute of Nuclear Physics, N.C.S.R.
``Demokritos'', GR-15310 Aghia Paraskevi, Attiki, Greece}
} 

\end{center}

\centerline{Abstract}
 
{\it 
A brief report of the topics which received attention during the discussion 
session II of the International Workshop on Symmetries and Low-Energy Phase 
Transitions in Nuclear-Structure Physics, held in Camerino on 9-11 October 
2005, is given. These include special solutions of the Bohr Hamiltonian
for various potentials, the study of triaxial shapes and of degrees of freedom 
other than the quadrupole one (octupole, scissors), as well as the search 
for experimental manifestations of the critical point symmetries E(5) 
and X(5), and of the recently proposed critical point supersymmetry E(5/4). 
}

\bigskip\noindent 
{\large\bf 1. Solutions of the Bohr Hamiltonian with different potentials}

\medskip
In the original E(5) \cite{IacE5} and X(5) \cite{IacX5} papers, an infinite 
square well potential in $\beta$ is used in accordance to the expectation 
that the potential at the critical point has to be flat and scale invariant. 

\medskip\noindent
{\bf 1.1 Solutions related to E(5)}

\medskip
Different potentials used in the E(5) framework include 
a well of finite depth \cite{Capriow}, 
the sextic oscillator \cite{sextic},
Coulomb-like and Kratzer-like potentials \cite{FortE5},
a linear potential \cite{Fortrev},  
Davidson potentials \cite{Dav,Elliott,Rowe} of the form 
$\beta^2 + \beta_0^4 /\beta^2$, where $\beta_0$ is the position of the 
minimum of the potential \cite{varPLB,varPRC}, 
the $\beta^4$ potential \cite{Ariasb4,Ariasb4b}, 
as well as the $\beta^6$ and $\beta^8$ potentials \cite{E5} 
(the $\beta^2$ potential \cite{Bohr,Wilets,Dussel} corresponding 
to the well known U(5) case \cite{IBM}). 
A hybrid model employing a harmonic oscillator for $L\leq 2$ and an 
infinite square well potential for $L\geq 4$ has also been developed
\cite{Raduta}.  
A recent review of this topic has been given in \cite{Fortrev}.  

L. Fortunato reported on a detailed study of the potential \cite{vanRoos}
$$ u(\beta) = {(1-\eta)\over 2} \beta^2 + {\eta\over 4} (1-\beta^2)^2,$$
which is known \cite{vanRoos} to correspond to the U(5)-O(6) transition region
of IBM, U(5) being obtained for $\eta=0$ and O(6) being obtained for 
$\eta=\infty$, the critical point occuring at $\eta_C=0.5$ and corresponding 
to a $\beta^4$ potential, in agreement with the findings of 
\cite{Ariasb4,Ariasb4b}. This potential has also been considered in 
\cite{IacE5,E5}, but only within the region $0\leq \eta \leq 1$. 
Fortunato allowed $\eta$ to obtain higher values, observing 
that best agreement between E(5) and the potential of Eq. (1) is 
obtained for $\eta\simeq 5$, in which case the potential develops a bump at 
the center. It was remarked that B(E2)s should be calculated before final 
conclusions can be drawn. It is however remarkable that a bump results  
in the X(5) framework when using an effective $\beta$ deformation, determined 
by variation after angular momentum projection and two-level mixing
\cite{LeviX5}, as well as in Nilsson-Strutinsky-BCS calculations 
\cite{ZhangSm} for $^{152}$Sm and $^{154}$Gd, which are good examples 
of X(5), as will be seen in subsec. 3.2~.  

\medskip\noindent 
{\bf 1.2 Solutions related to X(5)}

\medskip
Different potentials used in the X(5) framework include 
a potential with linear sloped walls \cite{Capriosl}, 
the confined $\beta$-soft (CBS) rotor model \cite{Pietr1,Pietr2} 
(which utilizes an infinite square well potential displaced from zero), 
Coulomb-like and Kratzer-like potentials \cite{FortX5}, 
Davidson potentials \cite{varPLB,varPRC}, 
as well as the $\beta^2$, $\beta^4$, $\beta^6$, and $\beta^8$ potentials 
\cite{X5}. 
A recent review of this topic has been given in \cite{Fortrev}. 

The special case in which $\gamma$ is frozen to $\gamma=0$, while 
an infinite square well potential is used in $\beta$, leads to an exactly 
separable three-dimensional model, which has been called X(3) \cite{X3}. 

The approximate separation of variables used in X(5) has been tested recently 
through exact numerical diagonalization of the Bohr Hamiltonian 
\cite{Caprio72}, using a recently introduced \cite{Rowe735,Rowe45,Rowe753}
computationally tractable version of the Bohr--Mottelson collective model.

\medskip\noindent  
{\bf 1.3 Triaxial solutions} 

\medskip
In the E(5) case \cite{IacE5} the potential is supposed to be 
$\gamma$-independent, while in the X(5) case \cite{IacX5} the potential
is supposed to be of the form $u(\beta)+u(\gamma)$, with $\gamma$ obtaining 
values close to $\gamma=0$. Another family of solutions of the Bohr 
Hamiltonian can be obtained by allowing the potential to be of the form 
$u(\beta)+u(\gamma)$ and confining $\gamma$ around $\gamma=30^{\rm o}$. 
An infinite square well potential in this case leads to the Z(5) solution 
\cite{Z5} if $\gamma$ is allowed to vary around the $\gamma=30^{\rm o}$ 
value (the $\gamma$-soft case), and to the Z(4) solution \cite{Z4} if 
$\gamma$ is fixed to this value, being a parameter and not a variable  
any more (the $\gamma$-rigid case). 
In the $\gamma$-soft case Coulomb-like and Kratzer-like potentials 
have been used \cite{Forttri}, while a solution similar to Z(5), but 
with a different $\gamma$-potential, has been given in \cite{Jolos}. 
A recent review of the various solutions of the Bohr Hamiltonian has been 
given in \cite{Fortrev}.

The characteristic differences among the various models were discussed 
in detail. It was pointed out that E(5) and Z(4) possess very similar ground 
state bands and $\beta_1$ bands, as well as similar intraband and interband 
B(E2)s obeying the same selection rules. The main difference between the two 
models occurs in the $\gamma_1$ band, which exhibits opposite odd-even 
staggering in the two models, since in E(5) its levels are exactly grouped as 
$2^+$, ($3^+$, $4^+$), ($5^+$, $6^+$), \dots, which is a feature of the 
underlying O(5) subalgebra, while in Z(4) the approximate grouping is 
($2^+$, $3^+$), ($4^+$, $5^+$), \dots, which is a feature of rigid triaxial 
models \cite{Davydov}. It is known \cite{Casten} that $\gamma$-soft and 
$\gamma$-rigid models provide very similar results if $\gamma_{rms}$ of 
the former equals $\gamma_{rigid}$ of the latter. Since no clear examples 
of triaxial nuclei have been identified, it is expected that Z(4) 
[and Z(5)] would provide results in reasonable agreement with experiment 
only when this condition is approximately fulfilled. 

The proton-neutron triaxiality occuring in the SU(3)$^*$ limit 
\cite{Dieperink} of IBM-2 was also discussed, in connection to relevant 
shape phase 
transitions occuring in the study of the phase structure of IBM-2 
\cite{AriasPRL,CaprioPRL,CaprioAP}. It was pointed out that the main 
features of proton-neutron triaxiality are a low-lying K=2 band and B(E2)s 
resembling very closely the predictions of the Davydov model \cite{Davydov}
with $\gamma=30^{\rm o}$. These findings are in agreement with the assumption 
of $\gamma\simeq 30^{\rm o}$ made in Z(5) and Z(4), as well as with 
the prediction of low-lying $\gamma_1$-bands in these models.    

For experimental manifestations of the transition towards triaxiality, 
$^{168,170}$Er were suggested as possible candidates. 

It was also pointed out that Z(4), as well as X(3), are solutions in which the 
separation of variables is exact. Therefore it might be easier to clarify 
their algebraic structure, while this task seems more difficult in the cases
of X(5) and Z(5), where separation of variables is approximate. 

\bigskip\noindent
{\large\bf 2. Additional degrees of freedom}

\medskip\noindent
{\bf 2.1 Octupole deformation}

\medskip
It is generally accepted that octupole deformation ($\beta_3$) has to be taken 
into account simultaneously with the quadrupole deformation ($\beta_2$), which 
plays a dominant role. In the approaches developed so far for the description 
of the transition from octupole vibrations to octupole deformation in the 
light actinides, either only axially symmetric shapes are taken into account 
\cite{AQOA}, or slight triaxiality is allowed \cite{Bizzoct}. 
In the Analytic Quadrupole Octupole Axially symmetric (AQOA) model \cite{AQOA}
the angles are left out from the very beginning and a single axis is used 
for both the quadrupole and the octupole deformations, with no relative 
angle in between. Both methods 
lead to a satisfactory description of $^{226}$Th and $^{226}$Ra 
involving an infinite square well potential \cite{AQOA,Bizzoct}. 
In these cases the $\beta_1$ bandhead is 
higher than the X(5) bandhead by a factor of almost two, in agreement 
with experiment. The question of a transition from octupole deformation 
to octupole vibrations in the rare earth region has also been raised,
along with suggestions to examine if the N=90 isotones $^{150}$Nd and
 $^{152}$Sm, which are known to be the best examples of X(5) 
(see subsec. 3.2), are also critical with respect to the transition from 
octupole deformation to octupole vibrations, a question which remains open. 

It was pointed out that one has to be very careful in separating out the 
intrinsic variables, as in the old work by Rohozi\'nski \cite{Rohozinski}. 
When building the spdf-IBM \cite{Engel1,Engel2} it became clear that 
separation of the intrinsic and angular variables was very difficult when 
taking into account only the $s$, $d$, and $f$ bosons, while it became easy 
when the $p$ boson was also included. The $p$ boson was also helpful for 
taking into account the center of mass motion. The detailed study 
of the phase space of spdf-IBM, in analogy to the similar study carried 
out recently for IBM-2 \cite{AriasPRL,CaprioPRL,CaprioAP} would be very 
helpful in clarifying shape phase transitions involving the octupole degree 
of freedom, but it is technically quite demanding \cite{Engel2}. 

It was also pointed out that one has to be very careful with the terminology, 
since octupole deformation (the merging of the ground state band and the 
nearby negative parity band into a single band with levels of alternating 
parity) occurs in nuclei which are near-vibrational or transitional
from the quadrupole point of view, while octupole
vibrations (negative parity bands built on bandheads corresponding to 
one or more quanta of octupole vibration, their levels thus lying 
systematically higher than the levels of the ground state band with similar 
angular momenta) are observed in nuclei which are well deformed from the 
quadrupole point of view.   

Furthermore, it was pointed out that there is no {\it a priori} reason for 
criticality in the transition from octupole deformation to octupole vibrations 
to occur in the same nuclei, in which criticality in the transition 
from quadrupole vibrations to axial quadrupole deformation occurs. 
Such a conclusion has to be based on experimental evidence, since the relevant 
models contain many drastic approximations, made in order to make them 
analytically soluble. 

\medskip\noindent
{\bf 2.2 Scissors mode}

\medskip
It is known \cite{Ziegler,Ranga,Herz} that there is a strong correlation 
between the low lying dipole excitations, referred to as the {\sl scissors 
mode} \cite{Richter} and quadrupole deformation. On the basis of systematics 
of experimental data in the Xe, Ba, Nd, Sm, and Gd isotopic chains it has been 
argued \cite{Scheck}
recently that the M1 scissors mode strength distribution exhibits 
a transitional behaviour in the same nuclei in which the transition from 
spherical to quadrupole deformed shapes is seen.
In other words, it is expected that the low-lying dipole mode would
exhibit critical behaviour in the same nuclei which are critical with 
respect to the quadrupole mode, since the former is driven by the latter, 
while no similar coincidence is expected {\it a priori} for the octupole 
mode, as already remarked. 
Recent experimental data on $^{124-136}$Xe \cite{Kneissl} reinforce this 
argument. 

No analytically soluble model describing a phase shape transition of the 
scissors mode has been constructed so far. The two-rotor model 
\cite{LoIudice,Palumbo} and its reformulation \cite{DeFran} could be 
an appropriate starting point for this effort. 

\bigskip\noindent
{\large\bf 3. Experimental manifestations of critical point symmetries 
and supersymmetries} 

\medskip\noindent 
{\bf 3.1 Experimental manifestations of E(5)}

\medskip
The first nucleus to be identified as exhibiting E(5) behaviour was 
$^{134}$Ba \cite{CZE5}, while $^{102}$Pd \cite{Zamfir} also seems 
to provide a very good candidate. Further studies on $^{134}$Ba \cite{AriasE2}
reinforced this conclusion. 
$^{104}$Ru \cite{Frank}, $^{108}$Pd
\cite{Zhang}, $^{114}$Cd \cite{Long}, and $^{130}$Xe \cite{Liu} 
have also been suggested as possible candidates. A systematic search
\cite{ClarkE5,Kirson} on available data on energy levels and B(E2) transition 
rates suggested $^{102}$Pd, $^{106,108}$Cd, $^{124}$Te, $^{128}$Xe, and 
$^{134}$Ba as possible candidates, singling out $^{128}$Xe as the best one, 
in addition to $^{134}$Ba.   

S. F. Ashley reported on recent measurements, performed at Yale of B(E2)s in 
$^{106}$Cd, and planned measurements of low-lying B(E2)s in $^{102-108}$Cd
at Cologne, since earlier B(E2) measurements, by Coulomb excitation, in 
$^{106,108}$Cd seem unreliable. Detailed information on B(E2)s will clarify 
the presence of E(5) examples in this region, suggested in \cite{ClarkE5}. 

S. Harissopulos reported on recent measurements carried out at Legnaro 
on B(E2)s of $^{102}$Pd \cite{Kalyva}, singling out this nucleus as the 
best example of E(5) so far, in agreement with earlier work \cite{Zamfir}, 
among other reasons because no backbending occurs in its ground state band, 
which remains in excellent agreement with the parameter-free E(5) predictions 
up to high angular momenta. 

S. Harissopulos also reported on measurements to be carried out in 
Jyv\"askyl\"a on $^{128}$Xe, which has been suggested \cite{ClarkE5} as 
a very good candidate of E(5), but also is not far from a Z(4) behaviour
\cite{Z4}. This is in agreement with a recent report \cite{Kneissl}
on measurements of E1 and M1 strengths of $^{124-136}$Xe carried out at 
Stuttgart, which provides evidence for a shape phase transition around 
$A\simeq130$. 

S. Harissopulos also mentioned, as possible E(5) candidates, $^{140}$Xe, 
which lies far from stability, and $^{48}$Ti, for which Coulomb 
excitation can be performed for the B(E2)s. 

\medskip\noindent 
{\bf 3.2 Experimental manifestations of X(5)} 

\medskip
The first nucleus to be identified as exhibiting X(5) behaviour was 
$^{152}$Sm \cite{CZX5}, followed by $^{150}$Nd \cite{Kruecken}. 
Further work on $^{152}$Sm \cite{ZamfirSm,Clark,CZK,Bijker} and $^{150}$Nd
\cite{Clark,CZK,Zhao} reinforced this conclusion. The neighbouring N=90 
isotones $^{154}$Gd \cite{Tonev,Dewald} and $^{156}$Dy \cite{Dewald,CaprioDy}
were also seen to provide good X(5) examples, the latter being of inferior
quality. In the heavier region, $^{162}$Yb \cite{McC162Yb} and 
$^{166}$Hf \cite{McC166Hf} have been considered as possible candidates.
A systematic study \cite{ClarkX5} 
of available experimental data on energy levels and B(E2) transition rates 
suggested $^{126}$Ba and $^{130}$Ce as possible good candidates, in 
addition to the N=90 isotones of Nd, Sm, Gd, and Dy. A similar study in 
lighter nuclei \cite{Brenner} suggested $^{76}$Sr, $^{78}$Sr and $^{80}$Zr 
as possible candidates.    

D. Balabanski reported on recent measurements carried out at Yale on 
$^{128}$Ce, the first results of which on the B(E2)s within the ground 
state band are promising, while the analysis is going on. R. F. Casten 
remarked that this is a nucleus with a $P$-factor \cite{Casten} of 4.8, 
which is close to 5, thus it is expected to be a very good candidate for X(5). 
F. Iachello remarked that $^{128}$Ce, having 8 valence protons and 12 valence 
neutron holes, matches $^{152}$Sm, possessing 12 valence protons and 8 
valence neutrons, thus it is expected to be a very good example of X(5), 
as $^{152}$Sm is. 

P. G. Bizzeti referred to the current situation in the lighter region.  
$^{122}$Ba \cite{Fransen} has been identified as a possible X(5) candidate, 
based on experimental information for the energy levels. However, 
preliminary B(E2) measurements performed in Legnaro indicate that 
the B(E2)s within the ground state band appear to be closer to the 
rotational values and not to the X(5) ones. A similar situation seems 
to occur in $^{124}$Ba, according to a recent analysis, communicated 
by A. Dewald, of an earlier measurement performed at Cologne.   
The energy levels of $^{104}$Mo also suggested a X(5) interpretation
\cite{Brenner,Bizzeti}, but B(E2) values turned out to favour the rotor 
interpretation \cite{Hutter}.

S. Harissopulos reported on recent GASP measurements on $^{178}$Os
\cite{DewaldJPG}, for which B(E2)s indicate that it is the first good example 
of X(5) in the $A\simeq 180$ region, and on $^{176}$Os \cite{DewaldJPG}, 
for which the analysis is still going on. 

\medskip\noindent
{\bf 3.3 Experimental manifestations of E(5/4)} 

\medskip
E(5/4) \cite{IacE54} corresponds to a particle with $j=3/2$ coupled to an 
E(5) core. Therefore possible candidates should be searched for among odd 
nuclei with the odd nucleon lying in a $j=3/2$ level. The first suspects 
were the neighbours of $^{134}$Ba, a good example of E(5). $^{135}$Ba 
and $^{133}$Ba have been measured at Yale. It is already clear that $^{135}$Ba 
is not an example of E(5/4) \cite{Fetea}, probably because of the presence of 
a $j=1/2$ level close to the $j=3/2$ one, while the analysis of $^{133}$Ba is 
ongoing. As pointed out by F. Iachello, mixing between the multiplets based 
on the 1/2 and 3/2 levels leads to raising of one of the multiplets and 
lowering of the other.  

S. Harissopulos pointed out that a multiplet in $^{129}$Xe, based on 
a 3/2 level lying slightly above the 1/2 ground state,  exhibits strong 
similarities to E(5/4). J. Jolie suggested that a U(6/20) \cite{U620}
supersymmetry [corresponding to single-particle orbits with $j=1/2$, 3/2, 
5/2, 7/2 coupled to an O(6) core]
might be more appropriate for an overall description of this nucleus, 
since it corresponds to a 1/2 ground state and has been found appropriate 
for describing nuclei in the A$\simeq$130 region \cite{U620}.  

F. Iachello suggested that the Ir-Au region near closed shells is maybe 
appropriate for looking for experimental manifestations of E(5/4), since 
the U(6/4) supersymmetry was found there \cite{IVI,FVI}. 
$^{63}$Cu, although of small size, 
could also be a candidate, as discussed in \cite{IacE54}. 

J. Jolie pointed out a recent study \cite{JolieU612} bringing together the 
concepts of supersymmetry and shape phase transitions through the use of an 
IBFM \cite{IVI,FVI} Hamiltonian including a vibrational term and a 
quadrupole-quadrupole interaction, using single-particle orbits 
with $j=1/2$, 3/2, 5/2, as in the framework of the U(6/12) supersymmetry 
\cite{IVI,FVI}. This approach has given good results in the Os-Hg region
\cite{JolieU612}. 

\bigskip\noindent
{\bf Acknowledgements}

\medskip
The author is grateful to S. F. Ashley, D. Balabanski, P. G. Bizzeti and 
L. Fortunato for providing written descriptions of their contributions.

\end{document}